\documentclass[aps]{revtex4}
\pdfoutput=1
\usepackage{graphicx}
\begin{document}

\title{Elastic properties of cubic crystals: Every's versus Blackman's diagram}
\author{T. Paszkiewicz \footnote{Author to whom correspondence should be addressed; electronic mail: tapasz@prz.edu.pl}, S. Wolski}

\begin{abstract}
{Blackman's diagram of two dimensionless ratios of elastic constants is frequently used to correlate elastic properties of cubic crystals with interatomic bondings. Every's diagram of a different set of two dimensionless variables was used by us for classification of various properties of such crystals. We compare these two ways of characterization of elastic properties of cubic materials and consider the description of various groups of materials, e.g. simple metals, oxides, and alkali halides. With exception of intermediate valent compounds, the correlation coefficients for Every's diagrams of various groups of materials are greater than for Blackaman's diagrams, revealing the existence of a linear relationship between two dimensionless Every's variables. Alignment of elements and compounds along lines of constant Poisson's ratio $\nu(\left\langle 100\right\rangle,\textbf{m})$, ($\textbf{m}$ arbitrary perpendicular to $\left\langle 100\right\rangle$) is observed. Division of the stability region in Blackman's diagram into region of complete auxetics, auxetics and non-auxetics is introduced. Correlations of a scaling and an acoustic anisotropy parameter are considered.} 
\end{abstract}.

\affiliation{Faculty of Mathematics and Applied Physics,\\
Rzesz{\'o}w University of Technology, \\ul. W. Pola 1, PL-35-959 Rzesz{\'o}w, Poland } 
\maketitle

\section{Introduction}
Recently, Ledbetter used Blackman's diagram to systematize physical properties and interatomic bonding of f.c.c. metals \cite{1}. According to this author, Blackman's diagram reveals among others the elastic anisotropy, proximity to Born mechanical instability, elastic-constants (interatomic-bonding) changes caused by alloying, pressure, temperature, phase transformations, and similarities in types of interatomic bonding. 

On the other hand, we used Everys' diagram for the unified description of elastic, acoustic, and transport properties of cubic media \cite{2}-\cite{8}. We also studied possible phases driven by mechanical instabilities of cubic crystals \cite{4}.  

In our note, we study relations between both types of diagrams, and apply them to various classes of elements and compounds. As a result we show that Every's diagrams reveal different characteristics of the elastic anisotropy than Blackman's .

\section{Geometrical relations between Blackman's and Every's diagrams} 
\label{sect:geom-relat}
The Blackman diagram is a plot of two dimensionless ratios of elastic constants
\begin{equation}
	F_{12}=\frac{C_{12}}{C_{11}},\;  F_{44}=\frac{C_{44}}{C_{11}},
	\label{eq:blackman-variables}
\end{equation}
where $C_{11}$, $C_{12}$, $C_{44}$ are three elastic constants characterizing cubic crystals. 

We used another set of parameters characterizing cubic elastic media introduced by Every \cite{9}. One of them has dimension of elastic constants  
\begin{equation}
	s_{1}=\left(C_{11}+2C_{44}\right),
	\label{eq:s1}
\end{equation}
and two remaining ones are dimensionless, 
\begin{equation}
s_{2}=\frac{\left(C_{11}-C_{44}\right)}{s_{1}}, \: s_{3}=\frac{\left(C_{11}-C_{12}-2C_{44}\right)}{s_{1}}. 
\label{eq:every-variables}
\end{equation}                              

The parameter $s_3$ characterizes acoustic and elastic anisotropies. It is equivalent to the familiar acoustic anisotropy parameter $K=C_{11}-C_{12}-2C_{44}$ \cite{9}. The parameter $s_{1}$ scales phase and group velocities of sound, Young's and shear moduli as well as surfaces of constant energy of long wavelength acoustic phonons.

The elastic anisotropy of a cubic crystal can be also characterized by the Zener anisotropy ratio $A$, which represents the ratio of the two extreme elastic-shear coefficients \cite{10}.
\begin{equation}
	A=2\frac{C_{44}}{C_{11}-C_{12}}. 
\label{Zener}
\end{equation}
In terms of Blackman's variables $A=\frac{2F_{44}}{1-F_{12}}$, whereas in terms of Every's variables $A=\frac{1-s_{2}}{3s_{3}/2-s_{2}+1}$. For isotropic media, $s_{3}=0$, and $A=1$. We shall note that at least two other sets of elastic parameters are used (cf. \cite{11} and \cite{12}). 

It is an easy task to express Every's variables in terms of Blackman's variables 
\begin{equation}
	s_{2}=\frac{1-F_{44}}{1+2F_{44}},\: s_{3}=\frac{1-F_{12}-2F_{44}}{1+2F_{44}}. 
	\label{eq:every-by-blackman}
\end{equation}

\begin{figure}[htbp]
	\centering
		\includegraphics{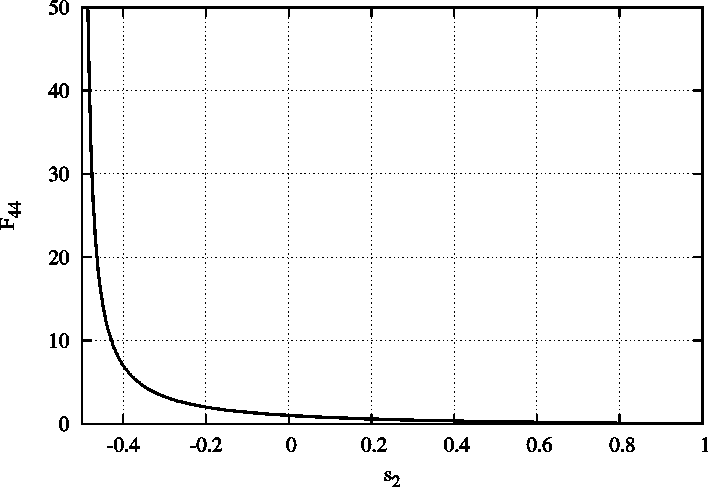}
	\caption{The dependence of $F_{44}$ Blackman's variable on $s_{2}$ Every's variable}
	\label{fig:f44}
\end{figure}

The inverse relations read
\begin{equation}
F_{44}=\frac{1-s_{2}}{2s_{2}+1},
F_{12}=\frac{4s_{2}-3s_{3}-1}{2s_{2}+1}.
\label{eq:blackman-by-every}
\end{equation}
The dependence of $F_{44}$ on $s_{2}$ is shown in Fig. \ref{fig:f44}.
Inspecting Fig. \ref{fig:f44}, we conclude that a very small portion of the ST gives rise to a large portion of the stability region in ($F_{12}$,$F_{44}$)-plane.

To a line 
\begin{equation}
	s_{3}=as_{2}+b,
	\label{eq:every-line}
\end{equation}
in ($s_{2},s_{3}$)-plane there corresponds a line 
\begin{equation}
F_{12}=(a-2b-2)F_{44}+(1-a-b),	
	\label{eq:blackman-line}
\end{equation}
in the ($F_{12},F_{44}$)-plane.  

\begin{figure}[htbp]
	\centering
		\includegraphics{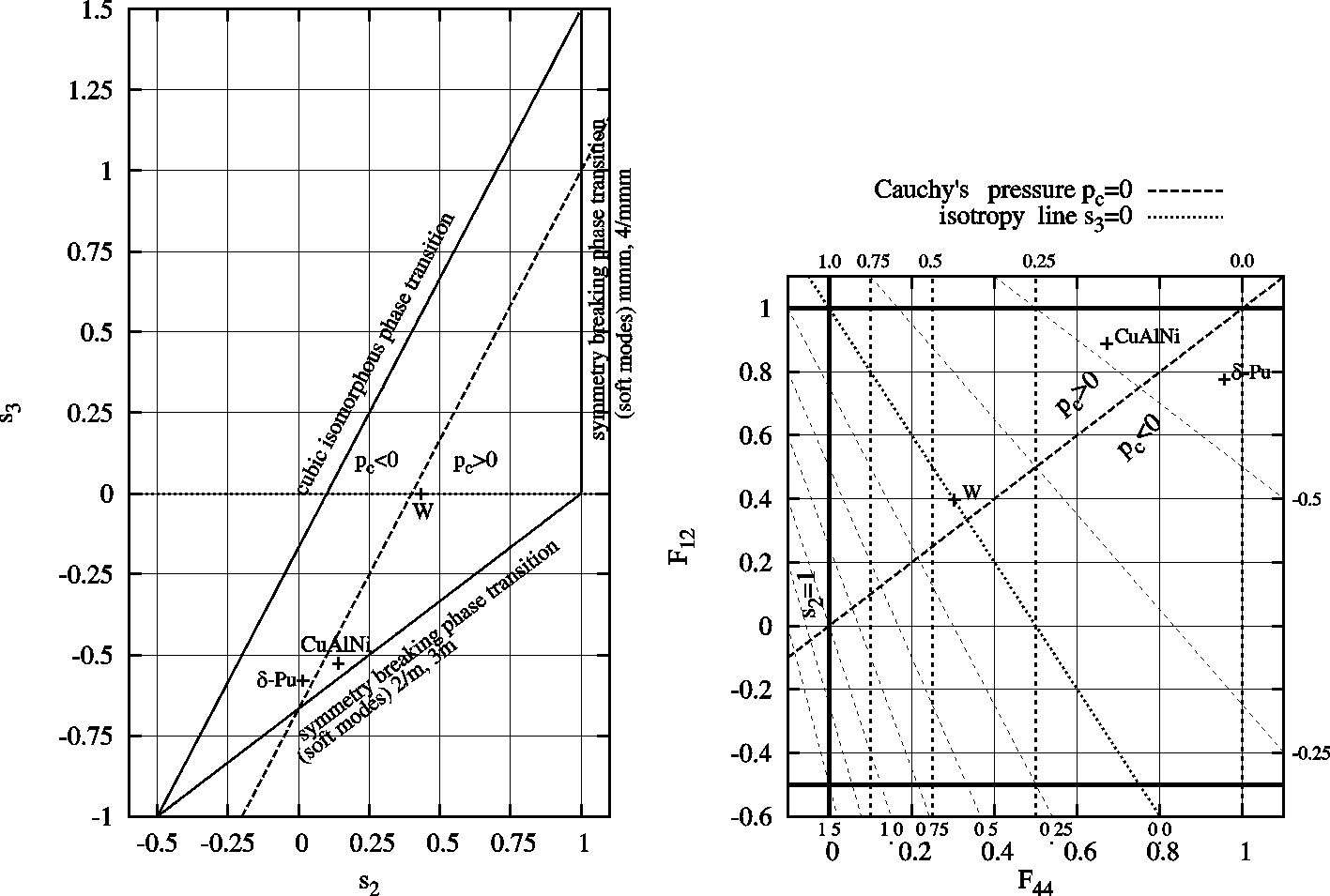}
	\caption{Characteristic lines in ($s_{2},s_{3}$)- and ($F_{12}, F_{44}$)-planes. Location of an isotropic element W and two strongly anisotropic materials $\delta$-Pu and CuAlNi shape memory alloy are indicated. The thick dashed lines represent the vanishing Cauchy's pressure.}
	\label{fig:char-lines-plot}
\end{figure}

The set of coordinates $s_{2}$, $s_{3}$ lets us partition a region in the ($s_{2},s_{3}$)-plane by means of vertical lines $s_{2}=s_{2}^{(i)}$ and horizontal lines $s_{3}=s_{3}^{(i)}$ (i=1,2,3\ldots,n). The corresponding level curves $F_{44}^{(i)}=\frac{1-s_{2}^{(i)}}{1+s_{2}^{(i)}}$ and $F_{12}^{(i)}=-2\left(s_{3}^{(i)}+1\right)F_{44}+\left(1-s_{2}^{(i)}\right)$ determine a corresponding partition of the region in the ($F_{12}, F_{44}$)-plane by thin dashed lines depicted in the right panel of Fig. \ref{fig:char-lines-plot}. We see that the transformation (\ref{eq:every-by-blackman}) deforms the regular mesh in the ($s_{2}$,$s_{3}$)-plane. 

Consider characteristic lines in ($s_{2},s_{3}$)-plane. To the isotropy line $s_{3}=0$ ($a=b=0$) there corresponds a line defined by equation $F_{12}=-2F_{44}+1$. The region of stability in ($s_{2},s_{3}$)-plane (ST for short) is bounded by lines having equations $s_{3}=5s_{2}/3-1/6$ (upper side of the stability triangle $a=5/3$, $b=-1/6$), $s_{2}=1$ (vertical side of ST, $F_{44}=0$) and $s_{3}=2s_{2}/3-2/3$ (lower side of ST, $a=2/3$, $b=-2/3$). To these lines there correspond straight lines $F_{12}=-1/2$, $F_{44}=0$, and $F_{12}=1$. We notice that in contrast to the mechanical stability region in ($s_{2},s_{3}$)-plane, the region of stability in ($F_{12},F_{44}$)-plane is not bounded. It extends along the horizontal axis from the origin of the coordinate system to $+\infty$. The regions of mechanical stability in both ($s_{2},s_{3}$)- and ($F_{12},F_{44}$)-planes are indicated in Fig. \ref{fig:char-lines-plot}.

The Cauchy condition $C_{12}=C_{44}$ is equivalent to relations $s_{3}=5s_{2}/3-2/3$, ($a=5/3,\: b=-2/3$) and $F_{12}=F_{44}$. This condition defines the zero Cauchy pressure ($p_{c}=0$). Region of the stability triangle for which $C_{12}>C_{44}$ corresponds to $p_{c}>0$, whereas for the opposite inequality $p_{c}<0$. The line of vanishing Cauchy's pressure and regions of positive and negative Cauchy's pressure in both ($s_{2},s_{3}$)- and ($F_{12},F_{44}$)-planes are depicted in Fig. \ref{fig:char-lines-plot}. 

\begin{figure}[htbp]
		\centering
		\includegraphics{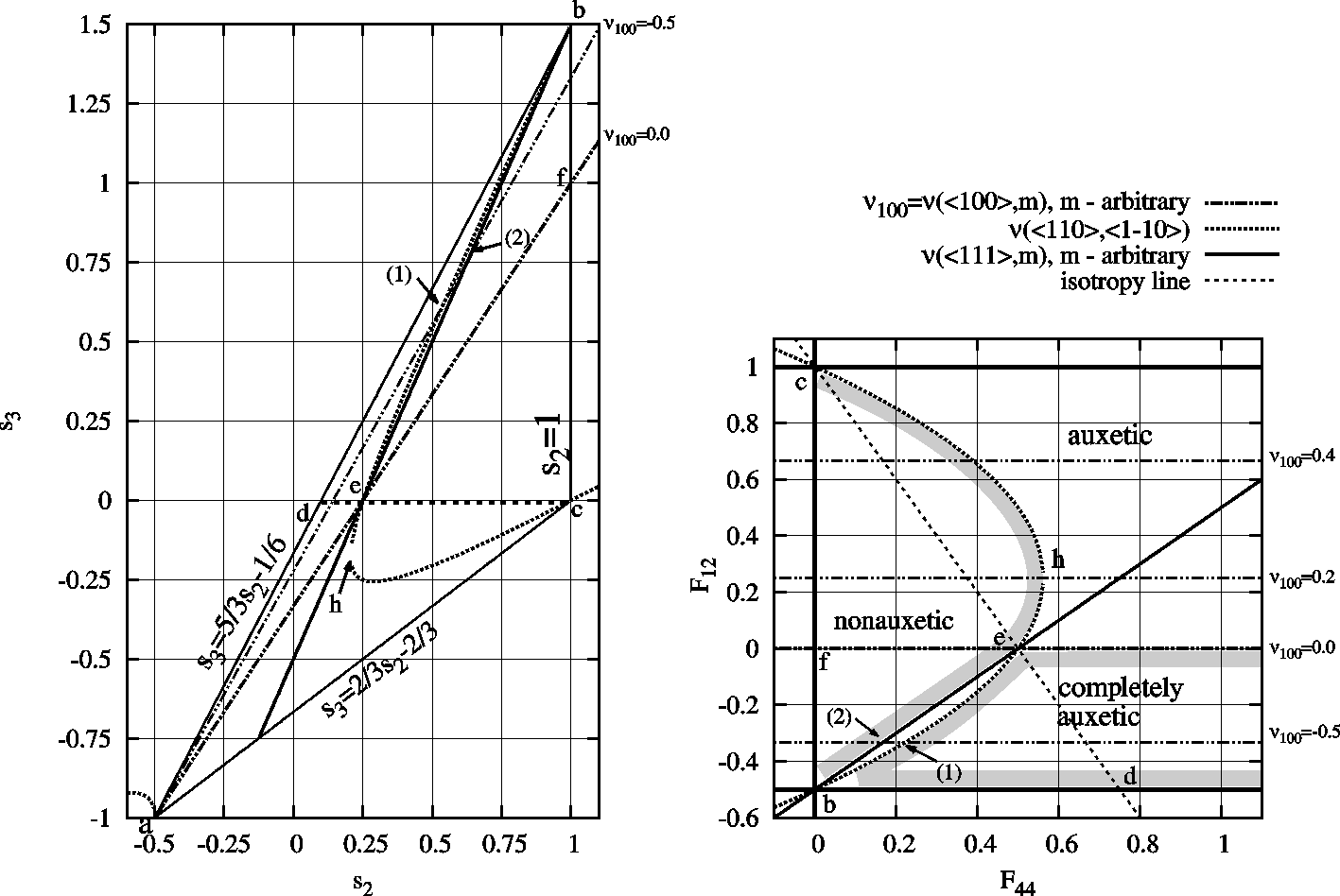}
		\caption{Auxetic properties of cubic elastic media -- characteristic regions in the stability triangle (left panel) and the stability region in ($F_{12}$,$F_{44}$)-plane (right panel). The full straight lines bound the stability region in this plane.}
	\label{fig:auxeticity-lines-plot}
\end{figure}

In our paper \cite{8} we introduced the division the stability triangle into regions of complete auxeticity, auxeticity and nonauxeticity. A crystal is completely auxetic if for every direction of longitudinal extension $\bf{n}$ and  perpendicular to it direction of the lateral contraction $\bf{m}$, Poisson's ratio $\nu(\bf{m},\bf{n})$ is negative. It is auxetic, if there exist pairs of directions $\bf{m}$ and $\bf{n}$ for which $\nu(\bf{m},\bf{n})<0$, and nonauxetic if $\nu(\bf{m},\bf{n})>0$ for each pair $\bf{m}$ and $\bf{n}$. The above regions are indicated on both Every's and Blackman's diagrams in Fig. \ref{fig:auxeticity-lines-plot}. Materials belonging to regions $ade$, and $deb$ of ST are completely auxetic. Materials belonging to the $ceh$ and $cfe$ regions are  nonauxetic. Materials belonging to the regions $aehc$ and $ebf$ are auxetics. This partition of the stability triangle generates the appropriate division of the stability region in the $(F_{12},F_{44})$-plane (cf. Fig \ref{fig:auxeticity-lines-plot}). 

In Fig. \ref{fig:auxeticity-lines-plot} we also draw lines of constant value of Poisson's ratio, which is an important characteristic of elastic media. In terms of $s_{2}$, and $s_{3}$, for $\textbf{n}=[1,0,0]$, ($\textbf{m}$-arbitrary) the suitable formula for $\nu_{100}=\nu([100],\textbf{m})$ ($\textbf{m}$ arbitrary) reads (cf. ref. \cite{7})
\begin{equation}
\nu_{100} = \frac{3s_{3}+1-4s_{2}}{3\left(s_{3}-2s_{2}\right)}.
\label{eq:n-in-s}
\end{equation}
 For a given value of $\nu_{100}$ Eq. (\ref{eq:n-in-s}) defines a straight line.

\section{Every's and Blackman's plots for cubic metals and compounds}
\label{sc:metals}

\begin{figure}[htbp]
	\centering
		\includegraphics{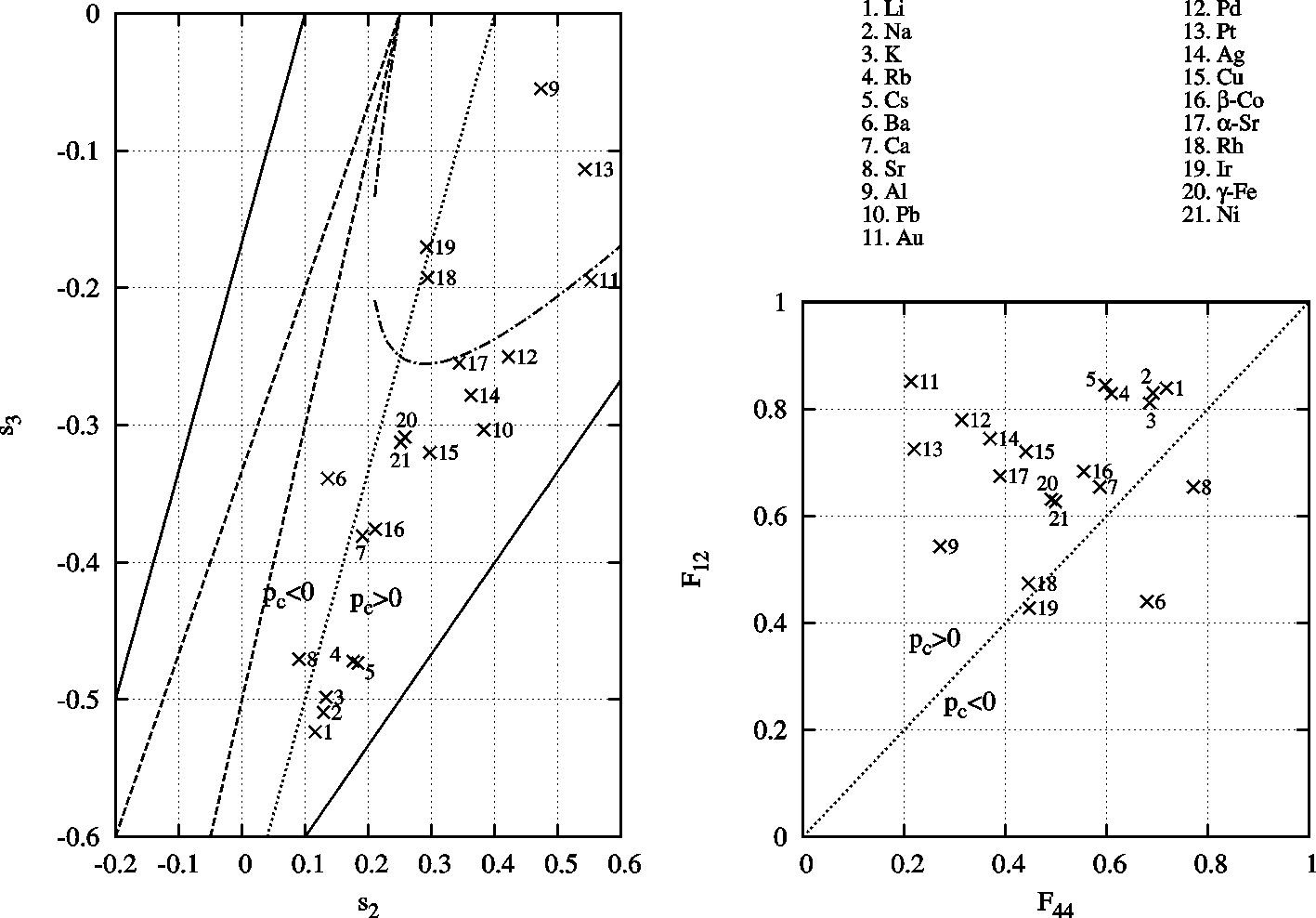}
		\caption{Every's and Blackman's diagrams for metals. Data for Li, Na, K, Rb, Cs, Ba, Ca, Sr, Al, Ga, Pb are taken form \cite{13}. For most of cubic metals, Cauchy's pressure is positive. The dotted lines indicate the vanishing Cauchy pressure.}
	\label{fig:metals}
\end{figure}

In Fig. \ref{fig:metals} we display Every's and Blackman diagrams for cubic metals. 

Inspecting both Every's and Blackman's diagrams, we notice that for most of metals Cauchy's pressure is positive. All depicted metals belong to the lower part ($s_{3}<0$) of the ST and most isotropic among the considered is Al, whereas the most anisotropic is $\delta$-Pu. For considered metals, the correlation coefficient $\rho_{E}$ for Every's plot  is equal to 0.86, whereas for Blackman's plot $\rho_{B}$ is smaller and equal to 0.03. One may expect that Every's plot reveals the existence of a factor which makes the relation among $s_{3}$ and $s_{2}$ almost linear. Most of cubic metals are auxetics, only Al, Ir, Pt and Rh are nonauxetic.  

Plutonium exists in six different crystallographic phases before it melts at the relatively low temperature of 640$^{\rm o}$ C. The two phases of greatest interest are the monoclinic $\alpha$-phase, the stable form of unalloyed plutonium at room temperature, and the face centered-cubic (fcc) $\delta$-phase, which can be retained down to room temperature by the addition of a few atomic percent of aluminum or gallium (e.g bout 0.6 $\%$ by weight of gallium \cite{14}). Pure $\delta$-plutonium, on lowering temperature transforms into the $\gamma$-phase which has a face centred orthorombic ($mmm$) structure. The high temperature fcc $\delta$-phase exhibits an unusual negative thermal expansion coefficient and has the largest low-temperature specific heat of any pure element. It is also the most elastically anisotropic fcc element. The shear moduli $C_{44}$ and $\left(C_{44}-C_{12}\right)/2$ differ by a factor of six, this is in contrast to that of normal metals, and is significantly higher than $\gamma$-cerium, lanthanum, and thorium. 

At low temperatures, the light alkali metals (Li, Na, K, Rb and Cs) undergo structural transformations from body-centered cubic to the far less symmetric close-packed rhomboedral ($mmm$) phase \cite{15}. 

With the exception of intermediate valent materials, the distributions of points on both kinds of diagrams differ. In Every's diagrams, they are elongated and characterized by correlation coefficients close to 1. In Blackaman's diagrams, their distributions are more random, hence the suitable correlation coefficients are rather small. In Table \ref{table1} we compare the correlation coefficients $\rho_{E}$ and $\rho_{B}$ for considered classes of materials. Since the correct value of the correlation coefficient depends on the number of points (i.e. number of accounted materials for each group), Table \ref{table1} contains such information.
\begin{table}[h]
\begin{tabular}{p{3cm} |p{2cm} p{2cm} p{2cm} p{2cm} p{2cm}}\cline{2-6}
                   & metals & oxides & intermediate valent & alkali halides & actinides-lanthanides\\
                   \cline{1-6}
No. of comp.   &21      &    18 & 8                 &     26  &        7            \\
 $\rho_{E}$        &0.86    & 0,73  &0.99               &   0.89      &        0.91         \\
 $\rho_{B}$        &0.03    & -0,13 &-0.93              &   0.37      &        0,42         \\
 \hline
 \end{tabular}
\caption{The values of correlation coefficient for various groups of cubic materials}
\label{table1}
\end{table}

\begin{figure}[htbp]
	\centering
		\includegraphics{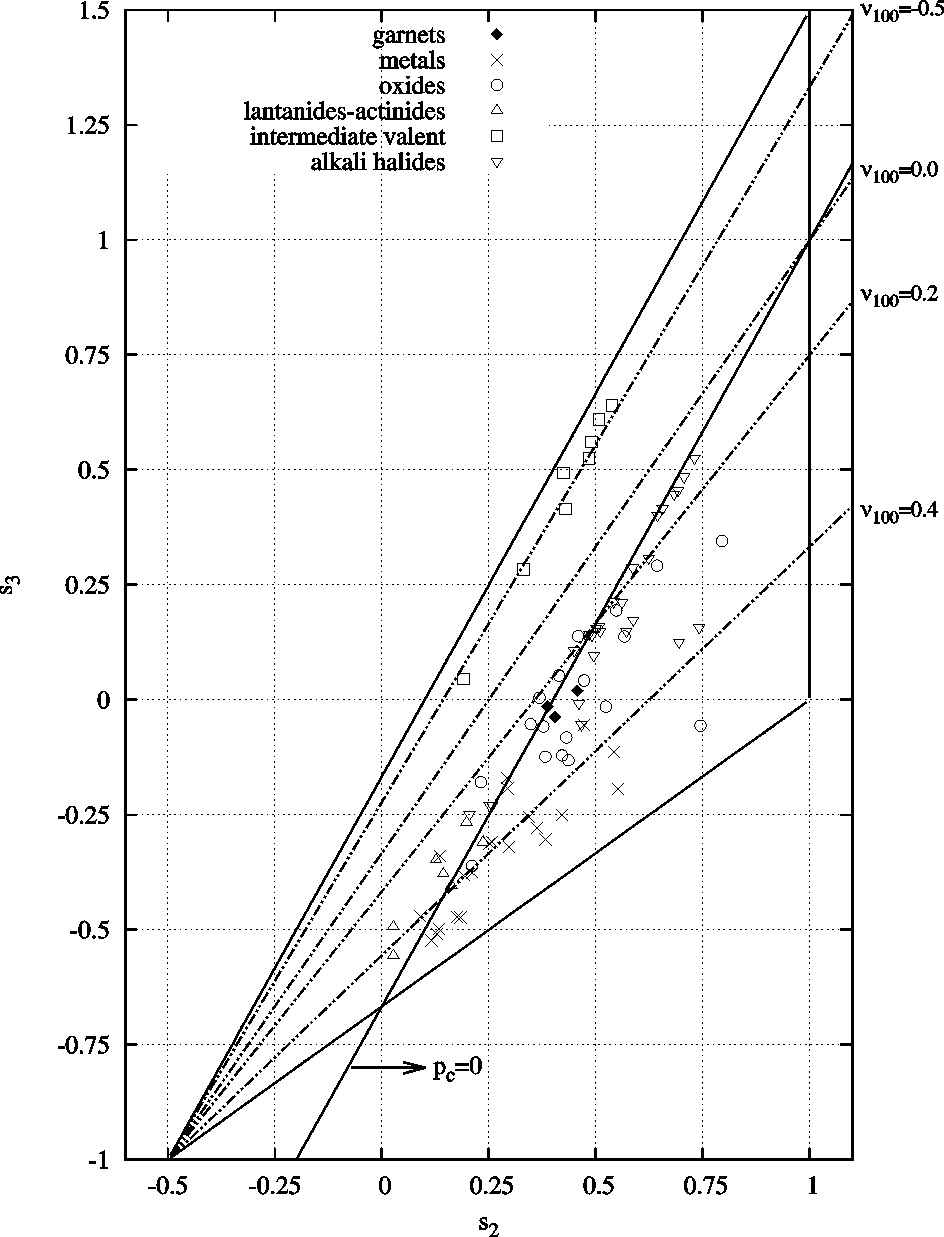}
	\caption{Every's diagram displaying all considered elements and compounds. Dashed lines correspond to constant values of Poisson's ratio $\nu_{100}$ (Eq. (\ref{eq:n-in-s}))}.
	\label{fig:all-plot-poiss}
\end{figure}
The Every's diagram, in which we display \textit{all} elements and compounds accounted by us, has a remarkable property (cf. Fig. \ref{fig:all-plot-poiss}). Namely, they have a tendency to be aligned along lines of constant values of Poisson's ratio $\nu_{100}$.

\begin{figure}[htbp]
	\centering
		\includegraphics{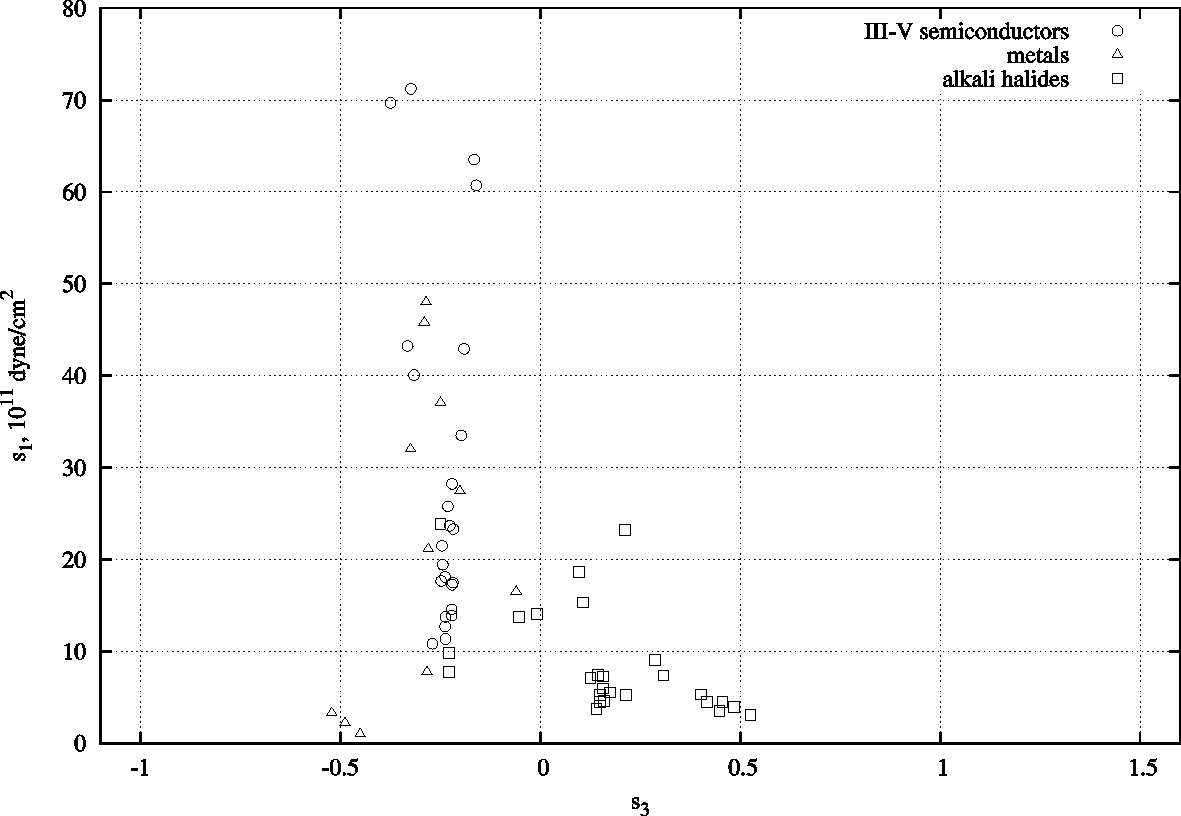}
	\caption{Correlations between parameters $s_{1}$ and $s_{3}$ for several classes of cubic elastic materials.}
	\label{fig:s1-s3-plot}
\end{figure}
In Fig. \ref{fig:s1-s3-plot} we show correlations between variables $s_{1}$ and $s_{3}$ (which is a measure of elastic anisotropy) for various classes of elastic cubic materials. In the case of III-V semiconductors,  values of $s_{1}$ ($10\leq s_{1}\leq 70$) are concentrated in a narrow interval around $s_{3}=-0.25$. In the case of metals, one distinguishes two groups of materials. One of them containing Li, Na and Cs, having rather small values of $s_{1}$, is located around $s_{3}\approx -0.5$ the other belongs to the narrow interval around $s_{3}\cong -0.25$ ($10\leq s_{1}\leq 50$). In the case of alkali halides, one marks also two groups. One of them is grouped around $s_{3}\approx -0.25$ ($8\leq s_{1}\leq 25$); the remaining belongs to the interval $0.1\leq s_{3}\leq 0.5$ ($2\leq s_{1}\leq 25$). All the above values of $s_{1}$ have to be multiplied by $10^{11}$ dyne/cm$^{2}$. 

\section{Properties of the most anisotropic element and compound -- $\delta$-Pu and CuAlNi shape memory alloy}
\label{sc:anisotropic}
Consider anisotropy properties of $\delta$-Pu and CuAlNi shape memory alloy. Values elastic constants of the austenite phase of CuAlNi ($C_{11}= 142.80$, $C_{12}= 126.84$ and $C_{44}= 95.90$ GPa) yield a very large value of Zener's anisotropy parameter $A = 12.02$ \cite{16}. The authors of ref. \cite{16} detected the onset of the stress-induced martensitic transformation by acoustic emission. In the case of $\delta$-Pu ($C_{11}= 34.56$, $C_{12}= 26.81$ and $C_{44}= 33.03$ GPa \cite{14}), thus $A=7.0$. These two elastic materials are representing by points lying in the left corner of ST, close to both -- upper and lower stability borders. 

Using these sets of elastic constants, we draw the slowness surfaces, the diagrams for Young's modulus, and dependence of Poisson's on direction of the lateral contraction $\bf{m}$ for a selected direction of longitudinal extension $\bf{n}$. 

\begin{figure}[htbp]
	\centering
		\includegraphics{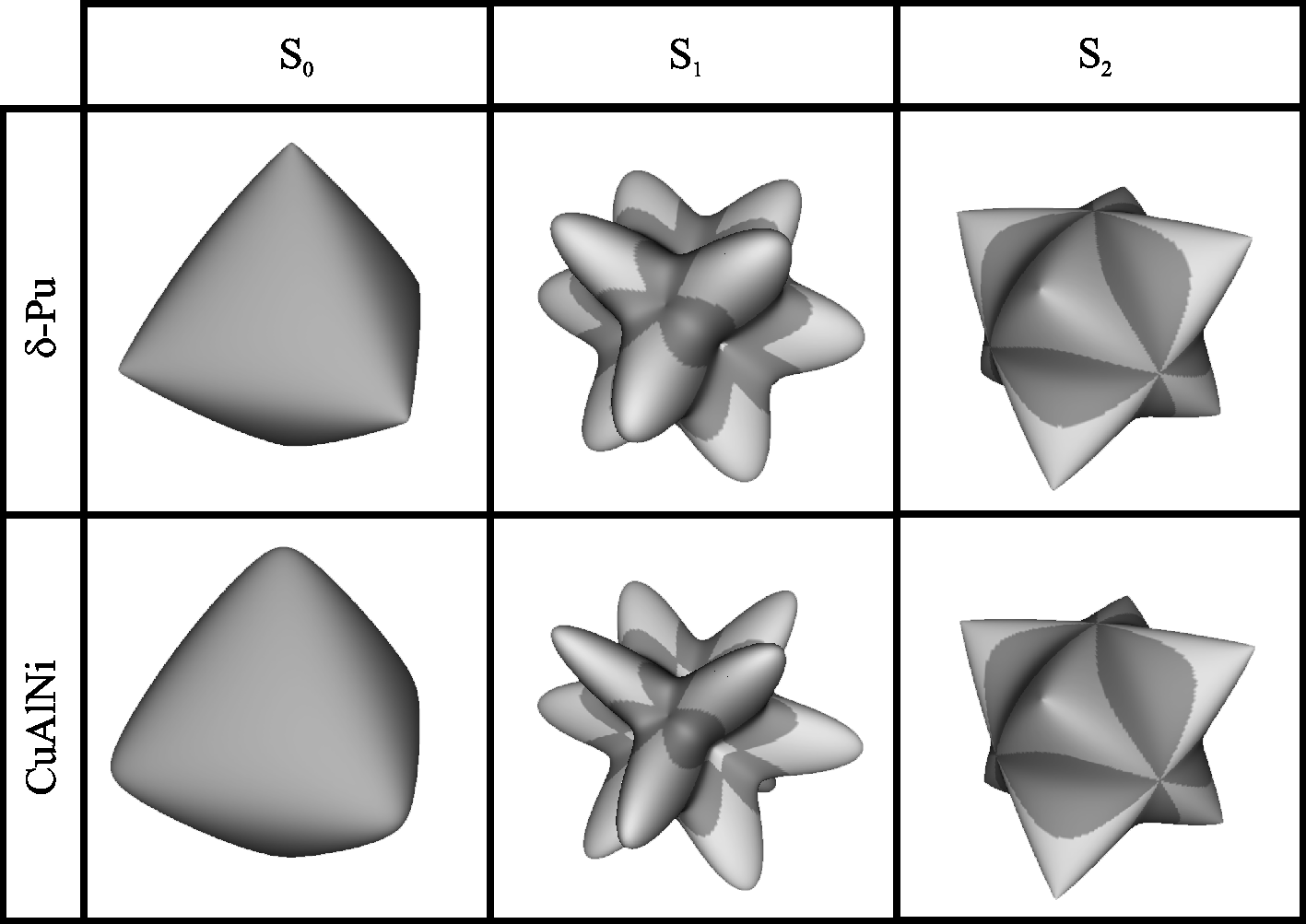}
	\caption{The slowness surfaces for strongly anisotropic materials -- $\delta$-Pu and CuAlNi shape memory alloy. For wolfram W the slowness surfaces are  spheres.}
	\label{fig:cualni-slowness}
\end{figure}

Similarly to the Fermi surfaces of metals, the slowness surfaces (s-surfaces) play a distinguished role in transport phenomena in gases of long-wavelength acoustic phonons (cf. \cite{6}). The branches of such phonons are indicated by the index $j=0,1,2$. The $j$-th slowness surface is similar to the constant energy surface for $j$-th branch of long-wavelength acoustic phonons. The $S_0$-surfaces for the fastest phonons are concave (cf. Fig. \ref{fig:cualni-slowness}).  Regions on $S_1$ and $S_2$-surfaces (respectively for the slowest and medium phase velocity phonons) where Gauss curvature $\Gamma$ is negative are shaded, whereas their regions with $\Gamma >0$ are brighter (Fig. \ref{fig:cualni-slowness}).    

\begin{figure}[htbp]
	\centering
		\includegraphics{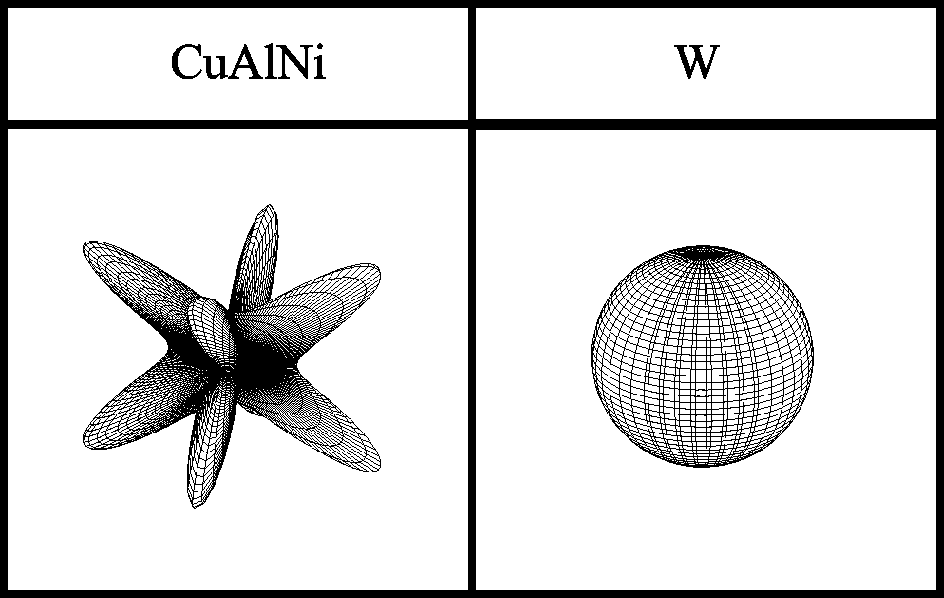}
	\caption{The angular dependence of Young's moduli of CuAlNi shape memory alloy and wolfram W. Maxima occur along $<111>$, minima along
$<100>$.}
	\label{fig:young}
\end{figure}
The diagrams \ref{fig:young} compare CuAlNi shape memory alloy, the most anisotropic known element, with wolfram -- an almost  isotropic element. CuAlNi's
Young's modulus shows eight lobes along the four $\left\langle 111\right\rangle$ directions with minima along the six $\left\langle 100\right\rangle$ directions. The contrast with wolfram is enormous. The maximal linear dimensions of both surfaces are normalized to unity. 

\begin{figure}[htbp]
	\centering
		\includegraphics{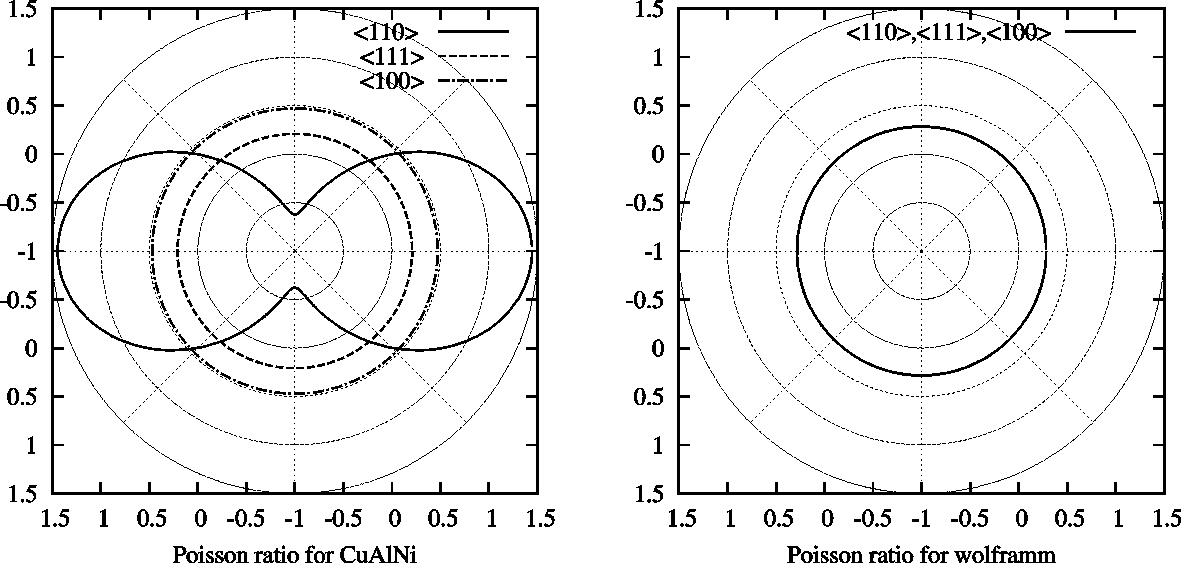}
	\caption{The angular dependence of Poisson's ratio of CuAlNi shape memory alloy and of wolfram W for extensional strains in the three crystalline directions $\left\langle 110\right\rangle$, $\left\langle 111\right\rangle$ and $\left\langle 100\right\rangle$.}
	\label{fig:poissonCuAlNi}
\end{figure}

For CuAlNi shape memory alloy, the Poisson ratio provides particular interest because it shows different profiles in different principal crystallographic
directions. In some directions, it shows large negative values (cf. left panel of Fig. \ref{fig:poissonCuAlNi}). A negative Poisson ratio means that pulling in one direction causes expansion in the transverse plane, a phenomenon associated traditionally with cork and network structures. 

Results for CuAlNi shape memory alloy are shown with those for wolfram in the illustrations at right. The illustrations (Fig. \ref{fig:poissonCuAlNi}) are for extensional strains in the three crystalline directions $\left\langle 110\right\rangle$, $\left\langle 111\right\rangle$ and $\left\langle 100\right\rangle$. For CuAlNi, for an extension in the $\left\langle 110\right\rangle$ direction, the lateral strain is highly anisotropic. Much current research proceeds on negative-Poisson-ratio materials, a material class called auxetics (cf. Sect. \ref{sect:geom-relat} and ref. \cite{17}and  \cite{18}). The diagrams for CuAlNi alloy dispel any notions of elastic similarity, isotropic behavior, and always-positive Poisson ratios.

For $\delta$-Pu diagrams for Young's modulus and for Poisson's ratio can be found in ref. \cite{19}. 


\end{document}